\documentclass[preprint,12pt]{elsarticle}
\usepackage{epsfig}
\usepackage{subfigure}
\usepackage{amssymb}
\graphicspath{{eps/}}
\usepackage{amsbsy}
\usepackage{amsmath}
\usepackage{color}
\usepackage{booktabs}
\usepackage{graphicx,adjustbox}

\usepackage{fancyhdr}

\begin{document}
\begin{frontmatter}

\address[lab1]{School of Astronautics, Beihang University, Beijing, 100083, China}
\address[lab2]{Shenyuan Honors College, Beihang University, Beijing, 100083, China}

\author[lab1,lab2]{Xinwei Wang}
\ead{wangxinwei@buaa.edu.cn}
\author[lab1]{Chao Han}
\ead{hanchao@buaa.edu.cn}
\author[lab1]{Rui Zhang}
\author[lab1]{Yi Gu}


\title{Scheduling multiple agile Earth observation satellites for oversubscribed targets using complex networks theory}

\begin{abstract}
The Earth observation satellites (EOSs) scheduling is of great importance to achieve efficient observation missions. The agile EOSs (AEOS) with stronger attitude maneuvering capacity can greatly improve observation efficiency while increasing scheduling complexity. The multiple AEOSs, oversubscribed targets scheduling problem with multiple observations are addressed, and the potential observation missions are modeled as nodes in the complex networks. To solve the problem, an improved feedback structured heuristic is designed by defining the node and target importance factors. On the basis of a real world Chinese AEOS constellation, simulation experiments are conducted to validate the heuristic's efficiency in comparison with a constructive algorithm and a structured genetic algorithm.
\end{abstract}


\end{frontmatter}

\thispagestyle{fancy}
 
\lfoot{\footnotesize \copyright~2019 IEEE. Personal use of this material is permitted. Permission from IEEE must be obtained for all other uses, in any current or future media, including reprinting/republishing this material for advertising orpromotional purposes, creating new collective works, for resale or redistribution to servers or lists, or reuse of any copyrighted component of this work in other works.} 
\cfoot{} 
\rfoot{}
\section{Introduction}
\label{sec:introduction}
 \setlength{\parindent}{2em}

Earth observation satellites (EOSs) equipped with unique cameras are specially designed to execute Earth observation missions. The number of the orbiting and plan-launching EOSs is increasing recently as a result of small-satellite technology development and lower satellite-launch costs~\cite{2014Nag}. Considering the advantages of expansive coverage area and long term surveillance, EOSs have been applied in the field of Earth resources exploration, natural disaster surveillance, and environmental monitoring. Therefore scheduling and management are of great importance for aerospace engineering~\cite{XiongLeus-27}, especially for multiple EOSs missions.

Previous literature characterizes the EOS scheduling problem into different mathematical models. On the basis of the engineering practice, Lin $et\  al$.~\cite{LinLiao-60} developed an integer programming model with salient features of sequence-dependent setup and job assembly. Constraint-satisfaction modeling was also introduced in the EOS scheduling~\cite{Bensana96exactand,Sun2008}. Gabrel $et\  al$.~\cite{GabrelMoulet-258} adopted graph theory concepts to describe the EOS scheduling problem, where the mission connection was transferred to directed acyclic edge. Similar works are conducted by Zufferey $et\  al$.~\cite{Zufferey2008} where the graph coloring techniques were utilized to develop the EOS scheduling model. Vasquez and Hao~\cite{VasquezHao-255} presented the problem as a generalized knapsack model, aiming at maximal profit function value while satisfying all kinds of constraints. Additionally, a window-constrained packing model was established in~\cite{WolfeSorensen-272}. Wang $et\  al$.~\cite{WangZhu-21,WangDemeulemeester-4} further extended the fundamental models by considering uncertainty of clouds and real-time scheduling.

To tackle various EOS scheduling models, considerable algorithms have been introduced and applied to realize effective scheduling schemes. According to the algorithm property, the existing algorithms for this study are classified into two categories: the exact and approximate ones. Exact algorithms are designed to achieve a global optimal solution. Gabrel and Vanderpooten~\cite{GabrelVanderpooten-61} solved the EOS scheduling problem under a multiple-criteria interactive procedure in a directed acyclic graph. Bensana $et\  al$.~\cite{Bensana96exactand} structured a depth first branch and bound algorithm on the basis of constraint satisfaction model. Benoist and Rottembourg~\cite{Benoist-104} introduced the Russian dolls approach to verify the upper bounds of the satellite scheduling problem with benchmark testing. Considering the EOS scheduling problem is NP-hard~\cite{WolfeSorensen-272}, the optimal solution is hardly trackable for large-scale EOS scheduling instances. Approximate algorithms are then adopted to approach a near optimal solution in a reasonable time frame. The intelligence algorithms including genetic algorithm~\cite{WolfeSorensen-272,BaekHan-162,KimChang-103}, local search algorithm~\cite{LemaitreVerfaillie-269} and ant colony optimization~\cite{ntagiou2018ant} have been widely applied for the EOS scheduling. Besides, tremendous heuristic procedures have been introduced to arrange feasible EOS scheduling missions. Based on a logic-constrained knapsack model, Vasquez and Hao~\cite{VasquezHao-255} developed a tabu search algorithm for daily scheduling of an EOS. The dynamic tabu tenure mechanism and techniques for constraint handling, intensification and diversification were testified on a set of large and realistic benchmark instances. Xu $et\  al$.~\cite{XuChen-5} employed priority-based indicators and sequential construction procedure to generate feasible solution. The algorithm performance was evaluated in various scenarios. Wolfe and Sorensen~\cite{WolfeSorensen-272} defined a fast and simple priority dispatch method to produce acceptable schedules. More heuristics works of EOS scheduling can be seen in~\cite{WangReinelt-50,chen2018priority}.

\begin{figure}[!h]
  \centering
  \includegraphics[width=4.0in]{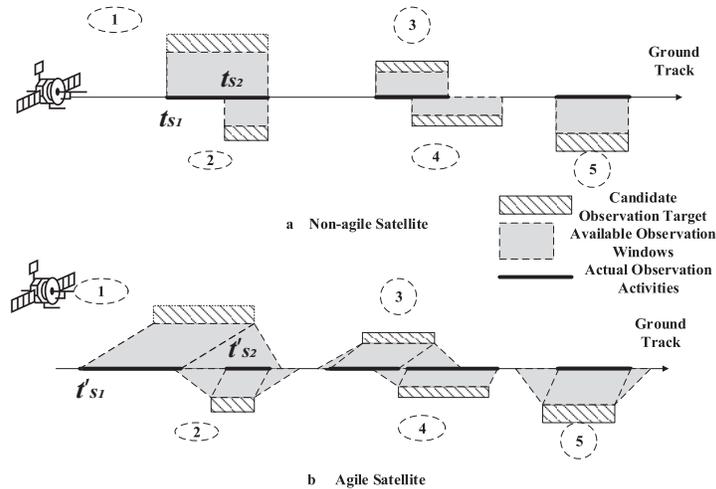}
  \caption{Different observation situations of agile and non-agile EOSs.}
  \label{fig:AEOS}
\end{figure}

Traditional non-agile EOSs equipped with cameras only have attitude adjustment ability along the roll axis, since the satellite platform is fixed in the direction of the pitch and yaw axes. As seen in Figure~\ref{fig:AEOS}(a), the non-agile satellite cannot start the observation process for target $1$ until the EOS arrives at $t_{s1}$. Different from traditional EOSs, the agile EOSs (AEOSs) with stronger attitude maneuver capability have freedoms along the roll, pitch and yaw axes. In Figure \ref{fig:AEOS}(b), the AEOS initializes the observation mission for target $1$ at $t_{s1}^{'}$ in advance, and begins the observation for target $2$ at $t_{s2^{'}}$ later than $t_{s2}$. The observation conflicts between target $1$ and $2$ have been solved. The AEOSs greatly improve observation efficiency, while the scheduling complexity is increased.

Lema\^{i}tre $et\  al$.~\cite{LemaitreVerfaillie-269} clearly defined the AEOS scheduling problem for the first time and proposed simplified versions of four different algorithms. Habet $et\  al$.~\cite{HabetVasquez52} formulated an AEOS scheduling as a constrained optimization problem, and considered stereoscopic and visibility constraints. Then a tabu search algorithm was designed with a systematic search using partial enumerations. Tangpattanakul $et\  al$.~\cite{TangpattanakulJozefowiez10} developed a multi-objective local search heuristic for an AEOS scheduling problem, where the proposed heuristic was compared to a biased random-key genetic algorithm. Aiming at a single area-target observation problem, Du $et\  al$.~\cite{du2018area} proposed a mission planning algorithm for a single AEOS taking the drift angle constraint of the imaging instrument into account. Valicka $et\  al$.~\cite{valicka2018mixed} introduced a novel deterministic mixed-integer programming model, and then extended it to a three-stage stochastic model with cloud cover uncertainty. Meanwhile, the theory of complex networks emerged, proving itself to be a valid tool in the domain of power distribution systems~\cite{DobsonCarreras-114}, Internet~\cite{Pastor-SatorrasVAzquez-119}, financial markets~\cite{IoriDeMasi-112}, economics~\cite{de2016topologic} and optimization systems~\cite{Liu-117}. Wang $et\  al$.~\cite{WangChen-7} modeled a single AEOS scheduling problem in complex networks, regarding each node as a discrete observation mission. Then a heuristic was proposed to obtain scheduling results effectively. Although practical constraints in the real world are not considered and multiple AEOSs cannot be handled in this model, quantitative insight into the multiple AEOSs scheduling problem by using complex networks knowledge could be obtained.

Multiple observation requirements for the same target are raised to fulfill stereo and time-series observation~\cite{GorneyEvans-107, NicholShaker-105, StearnsHamilton-106}. Under some special conditions, the target is supposed to be observed for several times in one or multiple orbits by one or even different AEOSs. However, none of the existing models and methods in this domain can be readily applied to address the multiple observation requirements. In this paper, each target on the ground is possibly desired to be observed more than once. The desired observation number for each target is designed as an input parameter according to user requirements, and the multiple-observation model is established in complex networks with attitude transformation, energy and memory capacity constraints. By defining factors of node and target importance, an improved feedback heuristic is proposed to solve the problem. The efficiency of the proposed heuristic is verified through comparisons with a constructive algorithm and a structured genetic algorithm. The performance of the feedback process is also tested.

The main contributions of this paper are threefold: (1) The multiple AEOSs, oversubscribed targets scheduling problem with multiple observations is addressed for the first time, where the satellite is capable of executing multiple observations in one orbit for the same target. (2) The scheduling problem with various constraints is modeled in the complex networks, and the potential observations are regarded as nodes in the networks. (3) An improved feedback heuristic is proposed by defining node and target importance factors.

The remainder of this paper is structured as follows. In Section 2 we describe the problem and establish mathematical model for multiple AEOS scheduling with constraints. The complex networks based heuristic with feedback is sketched in Section 3. The results including a series of computational experiments are reported in Section 4. We conclude the paper and point out future directions in Section 5.

\section{Model establishment}
\label{model}
\setlength{\parindent}{0pt}
\textbf{2.1 Problem statement}
\setlength{\parindent}{20pt}

This paper considers a multiple AEOSs, oversubscribed targets scheduling problem. Real-life AEOSs scheduling consists of satellite orbital operations, scheduling scheme upload and observation images download, making the problem too complicated to solve. To clearly state and simplify the problem, several necessary assumptions are listed as follows.

\begin{itemize}
  \item There are multiple AEOSs to execute observation missions. The oversubscribed candidate targets mean that the user requirements are already beyond the satellite observation capacity; as such, some candidate targets are abandoned.
  \item The satellite can only observe one target at a given time, and observation preemption is not allowed.
  \item The satellite is endowed with observation priority since each satellite has a different operational condition. Each target has an original profit and is possibly to be observed more than once. The desired observation number of each target is given as input.
  \item The requirements of the scheduling scheme upload and observation images download are not considered in the model, as it is assumed that there are enough ground data transmission stations to satisfy these requirements.
  \item The constraints of attitude transformation, energy consumption and memory capacity are introduced. Details of the constraints are described later.
\end{itemize}

Define $T$ as the set of the oversubscribed targets. For each target $i \in T$, the original profit is expressed as $\omega_{i}$ and $N_{i}$ is set as the desired observation number according to user requirements. Denote $S$ as the set of satellites. Considering that each satellite has different working conditions and observation cameras, the concept of observation priority is introduced and denoted as $\zeta_{j}$ for each satellite $j\in S$. Noting that one satellite could observe the target from different orbits, $O_{ij}$ is adapted to represent the orbit set of satellite $j$ for target $i$. Therefore the visible intervals can be expressed as $VI_{ijk}$ in each orbit $k \in O_{ij}$.

In accordance with to the complex networks theory, this paper intends to establish network nodes representing potential observation missions. To determine the specific observation time for each target, the continuous visible intervals $VI_{ijk}$ are further divided into several discrete observation windows (we refer to~\cite{GabrelMoulet-258} and \cite{WangChen-7} for further motivation of this choice). For instance, the visible interval is $VI_{ijk}=[100,150]$ where $100$ and $150$ are the interval beginning and ending time respectively. The observation duration for target $i$ is set as 5. Then $VI_{ijk}$ is divided into several observation windows as $[100,105], [110,115],..., [140,145]$. The index of discrete observation windows of $VI_{ijk}$ is denoted as $l$. The potential observation missions regarded as nodes in the networks now can be expressed as $OM_{ijkl}=[ts_{ijkl}, te_{ijkl}, \varpi_{ijkl}]$ where $ts_{ijkl}$ and $te_{ijkl}$ are specific observation beginning and ending times. $\varpi_{ijkl}$ is the mission profit defined as

  \begin{equation}
\varpi_{ijkl} = \omega_{i} \cdot \zeta_{j} \cdot cos(roll_{ijkl})
\cdot cos(pit_{ijkl})
 \end{equation}

 \setlength{\parindent}{0pt}
 where $roll_{ijkl}$ and $pit_{ijkl}$ are corresponding roll and pitch angles while executing the observation mission. The yaw angle is not related to the mission profit, since it does not significantly affect imaging quality.
\setlength{\parindent}{20pt}

The attitude transformation, energy consumption and memory capacity constraints are taken into consideration. The attitude maneuvering time $\Delta_{Mjk}(il,i^{'}l^{'})$ between $OM_{ijkl}$ and $OM_{i^{'}jkl^{'}}$ consists of attitude maneuvering time $\Delta_{Vjk}(il,i^{'}l^{'})$ and attitude stabling time $\Delta_{Sjk}(il,i^{'}l^{'})$. Considering the orbit period of the EOS lies between one to several hours, the attitude transformation constraint for two observation missions in different orbits are clearly satisfied. The transformation constraints are considered only on the condition that two observation missions are scheduled in the same orbit for the same satellite~\cite{GabrelVanderpooten-61}. The observation angles $roll_{ijkl}$ and $pit_{ijkl}$ are obtained by calculating position vectors of the satellite and target. Therefore the matrix elements of $\Delta_{Vjk}(il,i^{'}l^{'})$ and $\Delta_{Sjk}(il,i^{'}l^{'})$ are determined and set as input parameters.

The energy system is supported by the solar panel collecting energy from the Sun. To maintain satellite orbiting operations, the dynamic balance between the energy collection and consumption should be guaranteed within the scheduling horizon. Although the solar energy collection condition changes due to position variations from the Earth, Sun and satellite, the amount of energy collection in one orbit is near constant~\cite{wu2013two}. Therefore for satellite $j \in S$, the maximal energy capacity for satellite attitude maneuvering and camera imaging in one orbit is denoted as $Emax_j$. Two more input parameters of unit time imaging energy consumption and maneuvering energy consumption are defined as $Eui_j$ and $Eum_j$ respectively. Similarly, the satellite memory capacity in one orbit is defined as $Mem_j$ for $j\in S$ . The unit time imaging memory occupation for each satellite is assumed to be a constant and is denoted as $Sme_j$. Then the memory capacity constraint can be formulated in each orbit for each satellite.

\setlength{\parindent}{0pt}
\textbf{2.2 Mathematical formulations}
\setlength{\parindent}{20pt}

Different from the previous works related to the EOS/AEOS scheduling problem, this paper considers multiple observations for the same target. The binary decision variables are denoted as $x_{ijkl}$ for observation mission $OM_{ijkl}$, where $x_{ijkl}=1$ when $OM_{ijkl}$ is scheduled and $x_{ijkl}=0$ otherwise. On the basis of the problem statements and assumptions, the mathematical formulations are structured as

 \begin{flalign}
\text{maximize}   \sum\limits_{i\in{T}} \sum\limits_{j\in{S}}\sum\limits_{k\in{O_{ij}}}  \sum\limits_{l\in{VI_{ijk}}} \varpi_{ijkl} \cdot x_{ijkl}
\label{ObjFunc}
 \end{flalign}

subject to

 \begin{equation}
\sum\limits_{j\in{S}}\sum\limits_{k\in{O_{ij}}} \sum\limits_{l\in{VI_{ijk}}} x_{ijkl}\leq{N_i}, \forall{i}\in{T}  
\label{FreCon}
 \end{equation}

 \begin{equation}
 \begin{split}
x_{ijkl}\cdot x_{i^{'}jkl^{'}}(ts_{ijkl}-te_{i^{'}jkl^{'}}-\Delta_{Mjk}(il,i^{'}l^{'}))\Delta_{Bjk}(il,i^{'}l^{'})\geq 0, \\ \forall{l,l^{'}}\in{VI_{ijk}}, \forall{k}\in{O_{ij}}, \forall{j}\in{S}, \forall{i,i^{'}}\in{T} 
\label{TransCon}
\end{split}
 \end{equation}

 \begin{equation}
  \begin{split}
\sum\limits_{i\in{T}}\sum\limits_{l\in{VI_{ijk}}}  x_{ijkl}(te_{ijkl}-ts_{ijkl})\cdot Sme_{j}\leq{Mem_j}, \forall{k}\in{O_{ij}}, \\
\forall{k}\in{O_{ij}}, \forall{j}\in{S}, \forall{i}\in{T} 
\label{MemoryCon}
\end{split}
 \end{equation}

 \begin{equation}
  \begin{split}
\sum\limits_{i\in{T}} \sum\limits_{l\in{VI_{ijk}}} x_{ijkl}((te_{ijkl}-ts_{ijkl})\cdot Eui_{j} + \Delta_{Vjk}({il}^{p},{il})\cdot Eum_{j})\leq{Emax_j},\\ \forall{k}\in{O_{ij}}, \forall{j}\in{S}, \forall{i}\in{T}  
\label{EnergyCon}
\end{split}
 \end{equation}

 \begin{equation}
x_{ijkl} \in \{ 0, 1 \}, \forall{l}\in{VI_{ijk}},\forall{k}\in{O_{ij}}, \forall{j}\in{S}, \forall{i}\in{T}
\label{XCon}
 \end{equation}

The object function~\eqref{ObjFunc} aims to maximize observation profits with high-quality images. Observation constraints~\eqref{FreCon} ensure the number of scheduled observation missions does not exceed the desired observations, since redundant observations for the same target are not rewarded. In the transformation time constraints~\eqref{TransCon}, $\Delta_{Bjk}(il,i^{'}l^{'})$ are boolean variables. Before considering the mission transformation constraints, it is determined whether the sum of the ending time $te_{ijkl}$ of mission $OM_{{ijkl}}$ and attitude maneuvering time $\Delta_{Mjk}(il,i^{'}l^{'})$ is less than the beginning time $ts_{i^{'}jkl^{'}}$ of mission $OM_{{i^{'}jkl^{'}}}$. If this condition is satisfied, then $\Delta_{Bjk}(il,i^{'}l^{'})=1$ and $\Delta_{Bjk}(il,i^{'}l^{'})=0$ otherwise. By utilizing this parameter, meaningless transformation constraints can be avoided. Constraints~\eqref{MemoryCon} compute the amount of memory occupation for each orbit, and the scheduled observation missions cannot exceed corresponding satellite memory capacity. Constraints~\eqref{EnergyCon} restrict the energy consumption for each orbit, where ${ijkl}^{p}$ is the index of the precedent mission of $OM_{{ijkl}}$. When $OM_{{ijkl}}$ is the first scheduled mission in the orbit, the value of $\Delta_{Vjk}(il,{il}^{p})$ is equal to 0. The entire energy consumption of the scheduled missions is limited within one orbit.

\section{Solution approach}

By introducing the concept of complex networks, a fast approximate scheduling algorithm is designed for a single AEOS oversubscribed-targets problem without considering constraints and multiple observations~\cite{WangChen-7}. Inherited from this work, two important indicators are maintained and redefined for this paper: the node importance factor ($NIF$) and the target importance factor ($TIF$). A structured heuristic approach with a feedback process is then developed to deal with the multiple AEOSs scheduling problem with constraints and multiple observations.

\setlength{\parindent}{0pt}
\textbf{3.1 Node importance factor ($NIF$)}
\setlength{\parindent}{20pt}

Each potential observation mission $OM_{ijkl}$ is regarded as a node in complex works. To evaluate the importance of the node, the $NIF$ is supposed to address the influences in three parts as
 \begin{equation}
NIF_{ijkl} = NV_{ijkl} \cdot NC_{ijkl} \cdot NRW_{ijkl}
 \end{equation}
where $NV_{ijkl}$ stands for the observation mission node value, $NC_{ijkl}$ represents situations of node conflict, and $NRW_{ijkl}$ is defined as the node relative weight.

$NV_{ijkl}$ represents the comprehensive observation profit considering the current node as well as possible subsequent nodes in the same orbit. For the multiple AEOSs scheduling problem, the $NV_{ijkl}$ is defined as
  \begin{equation}
NV_{ijkl} =  \sum\limits_{i^{'}jkl^{'} \in NVS_{ijkl}} \frac{\varpi_{i^{'}jkl^{'}}}{ts_{i^{'}jkl^{'}}-te_{ijkl}-\Delta_{Mjk}(il,i^{'}l^{'})+10} + \varpi_{ijkl}
 \end{equation}
 where $NVS_{ijkl}$ is a set of observation missions that belong to the same orbit for mission $OM_{ijkl}$ and do not conflict with $OM_{ijkl}$.

The observation mission nodes cannot always be scheduled since the observation missions are restricted to the attitude maneuvering constraints. The node conflict situations are considered by introducing the definition of $NC_{ijkl}$ as
 \begin{equation}
 \begin{split}
NC_{ijkl} = 1/ln(\sum\limits_{i^{'}jkl^{'} \in NCS_{ijkl}} \frac{{NV_{i^{'}jkl^{'}}}}{nt_{i^{'}}} + e)
\end{split}
 \end{equation}
where $NCS_{ijkl}$ is a set of observation missions that belong to the same orbit for mission $OM_{ijkl}$ and conflict with $OM_{ijkl}$.

The node relative weight is defined to normalize $NIF$ for reasonable results. The $NRW_{ijkl}$ is expressed as
 \begin{equation}
NRW_{ijkl} = \frac{\varpi_{ijkl}}{\max \limits_{{i^{'}jkl^{'} \in NCS_{ijkl}}} \varpi_{{i^{'}jkl^{'}}}}
 \end{equation}
where $\max \limits_{{i^{'}jkl^{'} \in NCS_{ijkl}}} \varpi_{{i^{'}jkl^{'}}}$ stands for the maximal observation profit among the missions conflicting with $OM_{ijkl}$.

\setlength{\parindent}{0pt}
\textbf{3.2 Target importance factor ($TIF$)}
\setlength{\parindent}{20pt}

The $TIF$ is designed to concern the scheduling priority for targets. For target $i\in T$, its related nodes are ordered in descending $NIF$, and the top $N_{i}$ nodes are selected. If the number of nodes with target $i$ is less than $N_{i}$, all the nodes are picked. Although other nodes could be scheduled in the final scheduling schemes, the selected nodes can represent the target importance to some degree. Therefore the $TIF$ of target $i$ is expressed as

 \begin{equation}
 \label{TIF}
TIF_{i} = \frac{{\sum\limits_{ijkl\in TNS_{i}} {NIF_{ijkl}} }}{nt^{'}_{i}}
 \end{equation}
where set $TNS_{i}$ contains the indices of the top $nt^{'}_{i}$ nodes for target $i$ in descending $NIF$ order. $nt^{'}_{i}$ represents the smaller value between ${nt_{i}}$ and $N_{i}$, and $TIF_{i}=0$ when $nt^{'}_{i}=0$.

\setlength{\parindent}{0pt}
\textbf{3.3 Structured feedback heuristic($SFH$)}
\setlength{\parindent}{20pt}

By defining and calculating two indicators, $NIF$ and $TIF$, an improved structured feedback scheduling algorithm is proposed. In order to schedule the nodes (potential observation missions) with various constraints, $NIF$ is computed for each potential observation mission in the beginning. The values of $TIF$ are easily obtained according to Eq.~\eqref{TIF}. Then the whole nodes are divided into different groups in line with the corresponding targets. hte node groups are ordered with target scheduling priority $TIF$, and the nodes are ranked with the same target by descending $NIF$ order. Assume current scheduling target index as $i$, and check whether the current mission node $OM_{ijkl}$ satisfies all constraints in the mathematical model. If so, $OM_{ijkl}$ is added into $SSN_{j}$ which stands for the set of the scheduled observation missions of satellite $j$. Otherwise consider the next node for target $i$ until the whole nodes have been considered or the number of scheduled nodes for target $i$ equals $N_{i}$. The feedback process is activated when both of the following conditions are satisfied: the desired observation number of target $i$ is not fulfilled after considering all the nodes belonging to target $i$, and there still exist unscheduled nodes of target $i$.

\begin{figure}[ht]
  \includegraphics[width=5in]{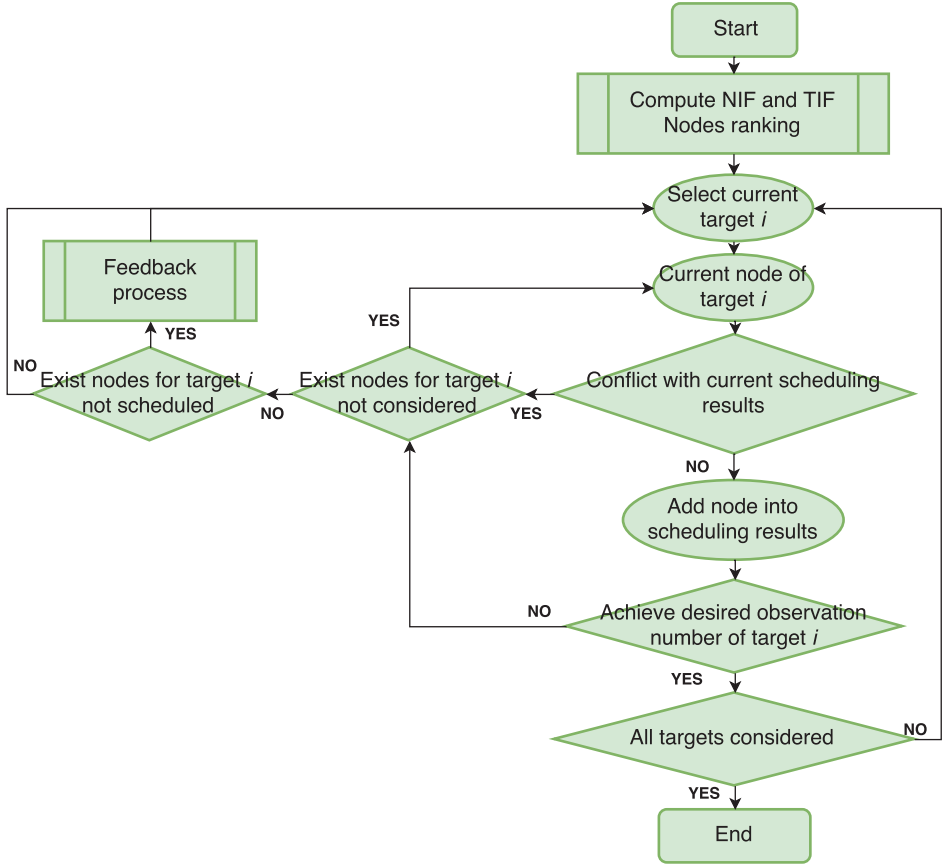}
  \caption{Flow chart of the feedback structured heuristic.}
  \label{fig:Diagram}
\end{figure}

The $SFH$ algorithm continues until all the targets are considered. Eventually, the scheduling solution for multiple AEOSs is obtained. The algorithm flow chart is shown in Figure \ref{fig:Diagram}.

Although the structure of $SFH$ is designed with comprehensive considerations, an optimal solution is hardly obtained. To narrow the gap between the optimal scheduling results and the solution obtained from the heuristic, a feedback process is proposed as follows:
\begin{itemize}
  \item Step 1: Rank the unscheduled nodes of target $i$ in descending $NIF$ order.
  \item Step 2: Mark the current unscheduled observation mission node $OM_{ijkl}$. If the energy and storage constraints are satisfied when adding $OM_{ijkl}$ into related $SSN_{j}$, go to Step 3; otherwise go to Step 4.
  \item Step 3: Check the amount of maneuvering conflicts with $SSN_{j}$. If $OM_{ijkl}$ conflicts with the mission related to the same target, or more than one scheduled mission nodes, consider the next unscheduled node of target $i$ and repeat Step 2; otherwise, temporarily remove the conflict-related mission node $OM_{i^{'}jkl^{'}}$ from the scheduled mission nodes set $SSN_{j}$ and proceed to Step 5.
  \item Step 4: If node $OM_{ijkl}$ has maneuvering conflicts with the nodes in $SSN_{j}$ or both of the energy and storage constraints cannot be satisfied when adding $OM_{ijkl}$ into $SSN_{j}$, consider the next unscheduled node of target $i$ and repeat Step 2; otherwise, mark subset of $SSN_{j}$ as $SSN_{j}^{sub-ijkl}$ for the mission nodes in orbit $k$ of satellite $j$. Then consider the nodes in the $SSN_{j}^{sub-ijkl}$ with ascending $TIF$ order. For the nodes belonging to the same target, the node with lower profit has higher priority. Remove the first node $OM_{i^{'}jkl^{'}}$ in $SSN_{j}^{sub-ijkl}$ from $SSN_{j}$ temporarily.
  \item Step 5: Check whether it is feasible to add $OM_{ijkl}$ and the unscheduled observation mission node $OM_{i^{'}j^{''}k^{''}l^{''}}$ of target $i^{'}$ in descending $NIF$ order into $SSN_{j}$. If the scheduled set ${SSN}^{'}_{j}$ is feasible and the inequality $\varpi_{i^{'}j^{''}k^{''}l^{''}} + \varpi_{ijkl} > \varpi_{i^{'}jkl^{'}}$ is satisfied, set $SSN_{j} = {SSN}^{'}_{j}$ and go to Step 6. Otherwise update mission node $OM_{i^{'}j^{''}k^{''}l^{''}}$ and repeat Step 5 until all the unscheduled nodes of target $i^{'}$ are considered. If $OM_{i^{'}jkl^{'}}$ belongs to the subset $SSN_{j}^{sub-ijkl}$, put $OM_{i^{'}jkl^{'}}$ into $SSN_{j}$, update $OM_{i^{'}jkl^{'}}$, remove the current mission node $OM_{i^{'}jkl^{'}}$ from $SSN_{j}$ temporarily and repeat Step 5 until all the nodes in $SSN_{j}^{sub-ijkl}$ have been considered. The removed node $OM_{i^{'}jkl^{'}}$ should be re-added into $SSN_{j}$ in the end.
  \item Step 6: If target $i$ has been scheduled with desired observation nodes or all of the unscheduled mission nodes of target $i$ have been considered, the feedback process ends; otherwise, consider the next node of target $i$ and go to Step 2.
\end{itemize}

\section{Experimental study}

In order to validate the proposed algorithm, various scenarios are designed on the basis of real world Chinese high resolution AEOSs $SuperView$. $SuperView$ is a commercial constellation of Chinese remote sensing satellites, of which four satellites have already been launched. The specific orbital parameters of the constellation are listed in Table \ref{tab:SatParameters}. The first column is the name of the satellite, and the parameters from columns 2 to 7 represent satellites' semi-major axis, inclination, right ascension of the ascending node, eccentricity, argument of perigee and mean anomaly respectively.

\begin{table}[htbp]
	\caption{Orbital parameters of the satellite constellation.}
	\centering
	\label{tab:SatParameters}
\begin{tabular}{ccccccc}
  \hline
  \toprule
    $ID$ & $a (km)$ & $i(^{\circ})$  & $\Omega(^{\circ})$ & $e$ & $\omega(^{\circ})$ & $M(^{\circ})$ \\
    \midrule
  $Sat1$ & 6903.673 & 97.5839 & 97.8446 & 0.0016546 & 50.5083 & 2.0288 \\
  $Sat2$ & 6903.730 & 97.5310 & 95.1761 & 0.0015583 & 52.2620 & 31.4501 \\
  $Sat3$ & 6909.065 & 97.5840 & 93.1999 & 0.0009966 & 254.4613 & 155.2256 \\
  $Sat4$ & 6898.602 & 97.5825 & 92.3563 & 0.0014595 & 276.7332 & 140.1878 \\
  \bottomrule
\end{tabular}
\end{table}

\begin{figure}[!ht]
  \centering
  \subfigure[Global distributions]
  {
    \includegraphics[width=3in]{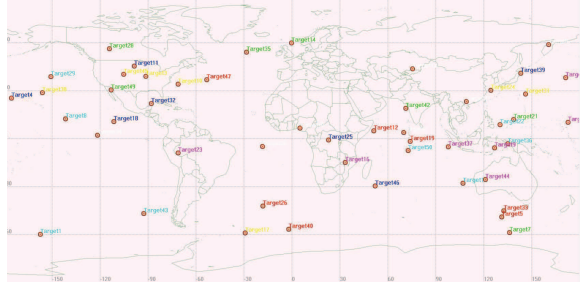}
  }
  \subfigure[Partial distributions]
  {
    \includegraphics[width=3in]{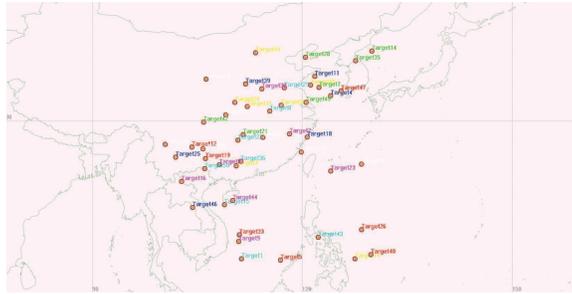}
  }
  \caption{Location distributions of the oversubscribed targets.}
  \label{fig:TarDistribu}
\end{figure}

The AEOSs of $SuperView$ are equipped with the same camera platform, and the agile platform allows up to $45^{\circ}$ maneuvers along the roll axis and $30^{\circ}$ maneuvers along the pitch axis. The observation priority of the satellite lies in $[5,10]$. As seen in Figure \ref{fig:TarDistribu}, the locations of the observation targets are assigned according to two categories: global uniform and partial centralized distributions. East Asia is selected as the partial distribution area. The original observation profit of the target is uniformly distributed in $[1,10]$ and the desired observation number for each target varies between $[1,5]$. The scheduling horizon is set as 24 hours with initial time as $1st$ January 2017, 00:00:00. The constraint parameters are described in Table \ref{tab:ConParameters}. Notice that the subscripts of constraint parameters are omitted, since all satellites share the same constraints parameters.

\begin{table}[!ht]
	\caption{Constraints parameters of the $SuperView$ satellites.}
	\centering
	\label{tab:ConParameters}
\begin{tabular}{cc}
  \hline
  \toprule
    Parameter & Value \\
    \midrule
  $Emax$ & 1000 \\
  $Eui$ & 5 \\
  $Eum$ & 10  \\
  $Mem$ & 1000 $Mb$ \\
  $Sme$ & 10 $Mb/s$ \\
  \bottomrule
\end{tabular}
\end{table}

The $SFH$ algorithm is implemented in C++ and tested on a laptop with Intel Core i5-7200U CPU (2.5GHz) under Windows 10 with 8 GB RAM. To verify algorithm effectiveness, the structured heuristic ($SH$) without feedback, a constructive first-in-first-out ($FIFO$) scheduling method~\cite{BianchessiRighini136} and a structured genetic algorithm are conducted. The general genetic algorithm with binary coding even has difficulty obtaining feasible solution, since the multiple observations and constraints for the same target are considered in the mathematical model. Therefore a similar algorithm process with the $SFH$ is constructed, by randomly initializing $NIF$ and $TIF$, and executing mutation and crossover rules during the iterations. The parameters and strategies of the genetic algorithm are given as follows:
\begin{itemize}
  \item Population size: 10
  \item Selection criterion: roulette wheel selection
  \item Crossover criterion: single point
  \item Mutation probability: 0.01
  \item Iteration: 500
  \item Maximal running time: 2000 seconds
\end{itemize}

The scheduling results of the global and partial distributions are reported in Tables~\ref{tab:ResultsGlobal} and \ref{tab:ResultsPartial}. In the first column, scenario remarks $G$ and $P$ stand for the global and partial distributions respectively, and the number that follows identifies the oversubscribed candidate observation targets. Columns $SFH$, $SH$, $FIFO$ and $SGA$ represent the objective function value obtained by corresponding algorithms. The unit of computation time in columns Time is seconds.

\begin{table}[htbp]
	\caption{Scheduling results of the global distributions.}
	\centering
	\label{tab:ResultsGlobal}
    \begin{adjustbox}{max width=\textwidth}
\begin{tabular}{cccccccccc}
  \hline
  \toprule
    Scenario & $SFH$ & Time & $SH$ & Time & $FIFO$& Time& $SGA$& Time \\
    \midrule
 $G\_50$ & 5915.1 & 0.08& 5893.3& 0.07& 4214.2& 0.13& 5151.5& 11.65 \\
  $G\_100$ & 10228.0 & 0.34& 10123.7& 0.27& 8532.6& 0.57& 9558.3& 52.09 \\
   $G\_150$ & 10572.0 &1.64 &10321.8 &0.56 &9539.2 &1.36 &9940.8 &52.69  \\
   $G\_200$ & 17168.0 &5.85 &16634.9 &1.10 &13790.4 &3.06 &14818.5 &207.84  \\
   $G\_250$ & 19875.9 &18.12 &18891.4 &1.14 &16873.5 &3.98 &17628.8 &533.59  \\
    $G\_300$ & 18784.1 &69.88 &17936.9 &1.85 &15361.6 &6.61 &15877.0 &1810.16  \\
  \bottomrule
\end{tabular}
\end{adjustbox}
\end{table}

\begin{table}[!ht]
	\caption{Scheduling results of the partial distributions.}
	\centering
	\label{tab:ResultsPartial}
\begin{adjustbox}{max width=\textwidth}
\begin{tabular}{cccccccccc}
  \hline
  \toprule
    Scenario & $SFH$ & Time & $SH$ & Time & $FIFO$& Time& $SGA$& Time \\
    \midrule
  $P\_50$ & 3027.0 &1.42 &2885.2 &0.07 &2284.1 &0.06 &2702.0 &11.22 \\
  $P\_100$  &3795.2 &6.59 &3411.2 &0.31 &2898.5 &0.32 &3032.7 &142.27 \\
   $P\_150$  &2966.9 &15.95 &2593.2 &0.66 &2104.3 &0.67 &2311.2 &465.84 \\
   $P\_200$  &3737.7 &22.64 &3273.4 &1.07 &2620.5 &1.19 &2893.9 &1072.15 \\
   $P\_250$  &4391.9 &42.79 &3991.5 &1.83 &2793.4 &2.01 &3261.5 &1843.11 \\
    $P\_300$  &4040.8 &63.83 &3597.9 &2.63 &2613.1 &2.91 &2912.6 &2000.00 \\
  \bottomrule
\end{tabular}
\end{adjustbox}
\end{table}

 It is observed in Table~\ref{tab:ResultsGlobal} that $SFH$ outperforms other methods, while the worst scheduling results are always obtained by $FIFO$ method. Although $SGA$ runs the largest computation time, the scheduling results are worse than the solutions of $SFH$ and $SH$. Note that $SH$ achieves better results than $FIFO$ and $SGA$ in the shortest time, and the gap between $SFH$ and $SH$ is slight especially for the small scale instances. Therefore, for scenarios with targets global distributions, the structured heuristic without feedback $SH$ is preferred when the fast scheduling requirement is raised.

As shown in Table \ref{tab:ResultsPartial}, $SFH$ achieves the best solution within an acceptable time frame considering target partial distributions. The scheduling results of $SH$ are significantly improved compared to the worst solution achieved by $FIFO$ method in the same running time. Similar to the global distributions, $SGA$ has the largest computation time and reaches maximal running time in Scenario $P\_300$.

\begin{figure}[!ht]
  \centering
 \includegraphics[width=5in]{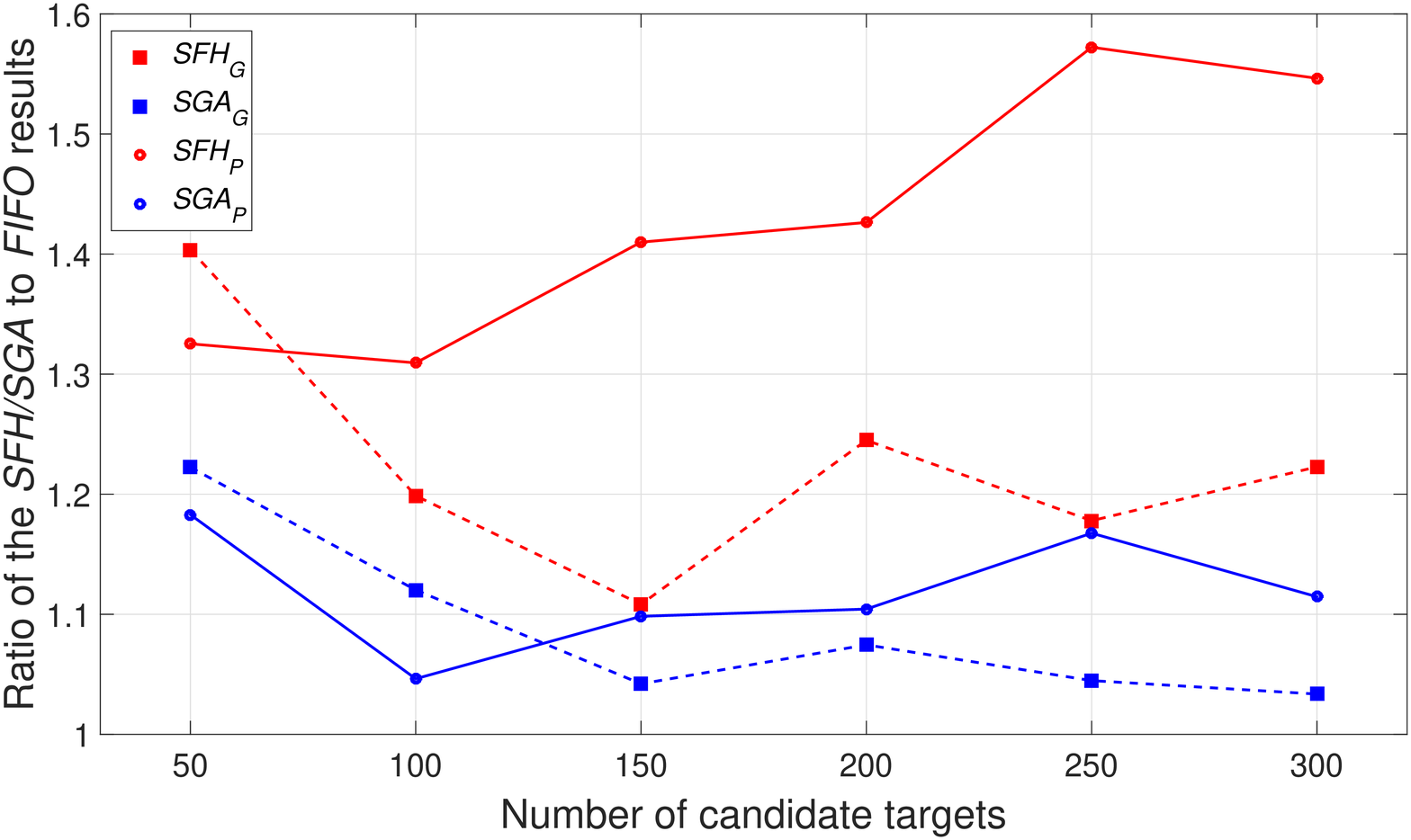}
  \caption{Solution comparisons.}
  \label{fig:RatioResultsFig}
\end{figure}

The comparisons between the $SFH/SGA$ to $FIFO$ results are described in Figure \ref{fig:RatioResultsFig}, where $SFH_{G}/SGA_{G}$ represent the ratio of $SFH/SGA$ to the $FIFO$ results in scenarios with target global distributions, and $SFH_{P}/SGA_{P}$ denote the ratio values in partial distribution situations. The scheduling results of $SFH$ increase 22.6\% in average compared to the results of $FIFO$ while $SGA$ achieves a 9.0\% improvement. The maximal increase of $SFH$ is observed in Scenario $G\_50$ since the satellites can effectively execute the high-profit observation missions. In the targets partial distribution scenarios, the performance of $SFH$ has a rising trend as the number of targets increases. The average profit addition of $SFH$ compared with $FIFO$ is 43.2\%. This is because oversubscribed targets in the partial distribution scenarios provide more opportunities to obtain a higher total observation profit. $SGA$ results raise 11.9\% compared with the results of $FIFO$, indicating that the performance of $SGA$ does not vary much from the target distributions and the number of targets.

\begin{figure}[!ht]
  \centering
  \subfigure[Feedback performance in different scenarios.]
  {
    \includegraphics[width=4in]{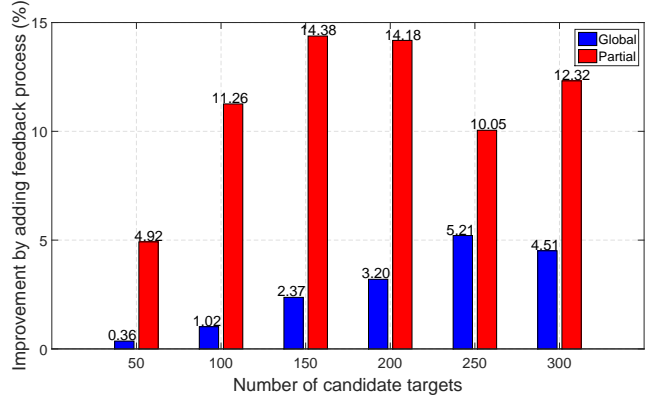}
  }
  \subfigure[Feedback working number in different scenarios.]
  {
    \includegraphics[width=4in]{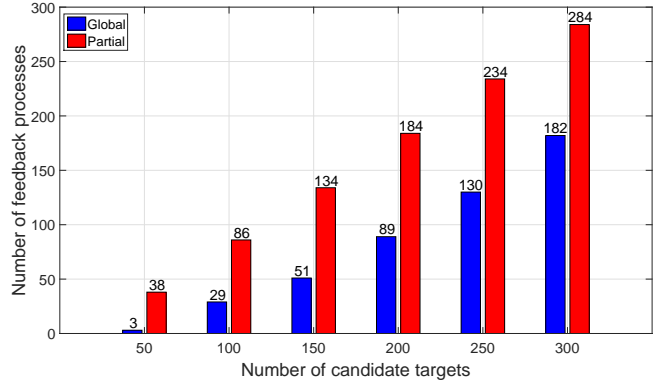}
  }
  \caption{Performance of feedback process.}
  \label{fig:FeedbackFig}
\end{figure}

To further distinguish the feedback performance in different scenarios, $SFH$ and $SH$ are compared in Figure \ref{fig:FeedbackFig}. According to Figure \ref{fig:FeedbackFig}(a), the $SFH$ only obtains 2.8\% improvement on average by adding a feedback process while the computation time of the $SFH$ increases exponentially. Different from the global distribution cases, the scheduling results of $SFH$ raise 11.2\% on average compared with the $SH$ results in the partial distribution scenarios. This is due to the different working numbers of the feedback process illustrated in Figure \ref{fig:FeedbackFig}(b). Clearly the numbers of the executed feedback process in the partial distribution scenarios are larger than that in global distribution cases. The more candidate observation targets are present, the more feedback processes are executed. For the partial distribution situations, the feedback process is triggered for more than 90\% of the targets overall.

This demonstrates that the feedback process is effectively applied in the partial distribution situations. Besides, the entire computation time is quite reasonable even on the large-scale scenarios with 300 targets, 4 satellites and 24-hour scheduling horizon. In conclusion, the $SFH$ outperforms other algorithms, especially for the multiple AEOSs, oversubscribed partial distribution targets scheduling problem.

\section{Conclusions and future directions}
\label{sec:conclusion}

The multiple AEOSs oversubscribed-targets scheduling with multiple observations problem is studied in this paper. To address this problem, the discrete observation windows are modeled as nodes in the complex networks at first. The factors of node and target importance are then defined to structure a feedback heuristic. The performance of $SFH$ has been verified in comparisons to $SH$, $FIFO$ and $SGA$ methods, and the efficiency of the feedback process is also validated in the experimental study. Overall, $SH$ without feedback performs well when fast mission scheduling is needed, and $SFH$ can obtain the best solution for all scenarios, especially for the instances with partially oversubscribed targets.

This research details how a complicated model can be combined with the complex networks theory to generate a structured feedback heuristic. Future potential directions include considering the data transmission process and continuous observation windows modeling. Data transmission is also a complicated scheduling problem and needs to be taken into consideration. Regarding continuous observation windows, future studies aim to determine the specified observation beginning and ending times by introducing interval scheduling models and algorithms.

\section*{Acknowledgment}
This research was supported by the China Scholarship Council and the Academic Excellence
Foundation of BUAA for PhD-students.

\bibliographystyle{IEEEtran}
\bibliography{SatScheduling}
\end{document}